\documentclass{PoS}

\usepackage{amssymb,amsmath,epsfig,color}
\usepackage{times,txfonts}
\usepackage{float} 
\usepackage[caption=false]{subfig} 

\def\to{\rightarrow}

\def\beq{\begin{equation}}
\def\eeq{\end{equation}}
\def\beeq{\begin{eqnarray}}
\def\eeeq{\end{eqnarray}}
\def\beal{\begin{align}}
\def\eeal{\end{align}}

\def\nn{\nonumber}
\def\b0{b_0}

\def\ID{1 \kern -.45 em 1}

\def\slash#1{\setbox0=\hbox{$#1$}               
        \dimen0=\wd0                            
        \setbox1=\hbox{/} \dimen1=\wd1          
        \ifdim\dimen0>\dimen1                   
        \rlap{\hbox to \dimen0{\hfil/\hfil}}    
        #1                                      
        \else              
        \rlap{\hbox to \dimen1{\hfil$#1$\hfil}} 
        /                                       
        \fi}                                    %

\title{NLO K-factors for Single-Inclusive Leptoproduction of Hadrons}

\ShortTitle{NLO K-factors for Single-Inclusive Leptoproduction of Hadrons}

\author{\speaker{Marc SCHLEGEL}\\
        University of Tuebingen\\
        E-mail: \email{marc.schlegel@uni-tuebingen.de}}

\author{Patriz HINDERER\\
        University of Tuebingen\\
        E-mail: \email{patriz.hinderer@uni-tuebingen.de}}

\author{Werner VOGELSANG\thanks{This work was supported by the ``Bundesministerium f\"{u}r Bildung und Forschung'' 
(BMBF) grant 05P12VTCTG. }\\
        University of Tuebingen\\
        E-mail: \email{werner.vogelsang@uni-tuebingen.de}}

\abstract{In these proceedings we discuss next-to-leading order (NLO) perturbative-QCD calculations of the cross section for 
the process $\ell N\to h X$. We briefly present the main features of the calculation and in particular analyze the role of quasi-real photons that enter the processes because the scattered lepton is 
not observed. We find that the NLO corrections are sizable for the spin-averaged cross section. In order to quantify this statement we present the K-factors of the cross sections for fixed target experiments HERMES, JLab12 and COMPASS.}

\FullConference{QCD Evolution 2015,\\
		26-30 May 2015\\
		Jefferson Lab (JLAB), Newport News Virginia, USA}

\begin{document}

\section{Introduction}

The process $\ell N\to h X$, i.e., the single inclusive production of a hadron at large 
transverse momentum in lepton-nucleon scattering, has attracted a lot of interest recently, both experimentally~\cite{E155,HERMES,Hulse:2015caa,JLab} and 
theoretically~\cite{Koike:2002gm,Kang:2011jw,Gamberg:2014eia,GPM1,GPM2,GPM3}. 
The reason for the interest in $\ell N\to h X$
comes from the study of single transverse-spin phenomena in hadronic scattering processes. It is well known that large
single-spin asymmetries have been observed~\cite{Aidala:2012mv} for the process $pp^\uparrow \to hX$, where
$p^\uparrow$ denotes a transversely polarized proton. To explain the large size of the asymmetries, and
their persistence all the way from fixed-target to collider energies, has posed a major challenge to theory. 
Although a lot has been learned, it is fair to say that a fully satisfactory understanding has yet
to be obtained. Measurements of corresponding asymmetries in the kinematically equivalent, but much simpler, 
processes $\ell N^\uparrow\to h X$, $\ell N^\uparrow\to \mathrm{jet}\, X$ have the promise to shed new light on the 
mechanisms for single-spin asymmetries in QCD. First fairly precise experimental data for 
$\ell N^\uparrow\to h X$ have recently been released by the HERMES~\cite{HERMES,Hulse:2015caa} and 
Jefferson Lab Hall A~\cite{JLab} collaborations.

We note that at first sight one might consider the related process $\ell N^\uparrow \to \ell^\prime X$ (which is just 
the standard inclusive deep-inelastic (DIS) process) to be equally suited for transverse-spin studies
in lepton scattering. However, the analysis of the corresponding single-spin asymmetry is considerably 
more complex because higher order QED effects are required for the asymmetry to be 
non-vanishing~\cite{Christ:1966zz,TPE1,TPE2,TPE4,TPE5}. 
In the same spirit as $\ell N^\uparrow\to h X$, also the processes $\vec{\ell} N^\uparrow\to h X$~\cite{DSA} 
with longitudinal polarization of  the lepton and $\ell N\to \Lambda^\uparrow X$~\cite{Lambda} with a transversely 
polarized $\Lambda$ hyperon have been considered in the literature recently. 

The proven method for analyzing single-inclusive processes such as $pp\to hX$ or $\ell N\to h X$ 
at large transverse momentum rests on QCD perturbation theory and collinear factorization. 
For single-transverse-spin observables, this involves a twist-3 formalism in terms of three-parton
correlation functions of the nucleon or the fragmentation 
process~\cite{QS1,Kanazawa:2000hz,Kouvaris:2006zy,Kang:2010zzb,Kanazawa:2011bg,Beppu:2013uda,FF1,FF2,FF3,pp2piX}. 
Interestingly, the recent study~\cite{pp2piX} suggests that the twist-3 fragmentation effects 
could be the dominant source of the observed large transverse-spin asymmetries in $pp^\uparrow \to hX$.

The collinear twist-3 approach has been used recently to obtain 
predictions for the spin asymmetry in $\ell N^\uparrow\to h X$. In Ref.~\cite{Gamberg:2014eia} a 
leading order (LO) twist-3 analysis has been presented in terms of parton correlation functions that were
previously extracted from data for $pp^\uparrow \to hX$. The results obtained in this way fail to 
describe the HERMES data~\cite{HERMES,Hulse:2015caa} for the spin asymmetries in 
$\ell N^\uparrow\to h X$. A comparison of perturbative calculations to the corresponding JLab  
data~\cite{JLab} is not possible as the data are for hadrons with transverse momenta below 1~GeV.

In our view it is premature to draw any conclusions from these findings at LO. Given the kinematics
(and the precision) of the present data, one may expect higher-order QCD corrections to the cross 
sections and the asymmetry to be important~\cite{Gamberg:2014eia} for a meaningful comparison of 
data and theory. At least next-to-leading order (NLO) corrections should be included. We stress that 
the twist-3 formalism, although so far only developed to LO, offers a well-defined framework for a 
perturbative study of the transverse-spin asymmetry in $\ell N^\uparrow\to h X$.

In a recent paper~\cite{ourpaper}, we took a first step toward an NLO calculation of the transverse-spin asymmetry for
$\ell N^\uparrow\to h X$ by computing the NLO corrections to the spin-averaged cross section for the process,
which constitutes the denominator of the spin asymmetry. In the following we will briefly present the main steps of the NLO calculation of Ref.~\cite{ourpaper}. Numerical predictions of the unpolarized cross-sections for several future and present-day experiments were presented as well in~\cite{ourpaper}. We found in~\cite{ourpaper} that the NLO corrections can become quite sizeable in particular for experiments at lower energies. It is illustrative to consider so-called K-factors, i.e. the ratio of the NLO-result divided by the LO-result for a certain observable. The K-factors quantify the size of the NLO-correction, and we will give numerical predictions for those K-factors in these proceedings.


\section{NLO calculation \label{nlocalc}}

First, we give a brief review of the calculation presented in Ref.~\cite{ourpaper} of the spin-averged cross section of the process $\ell(l) + N(P)\rightarrow h (P_h)+X$ up to NLO accuracy in pQCD. A large transverse momentum $P_{h\perp}\gg \Lambda_{\mathrm{QCD}}$ of the produced hadron sets a hard scale, so that 
perturbative methods may be used for treating the cross sections. It is useful to introduce the Mandelstam variables 
as $S=(P+l)^2$, $T=(P-P_h)^2$ and $U=(l-P_h)^2$. Furthermore, we label the energy of the detected hadron as 
$E_h$ and its three-momentum by $\vec{P}_h$. 

In collinear leading-twist perturbative 
QCD the hadronic cross section is approximated by convolutions of hard partonic scattering 
cross sections and parton distribution/fragmentation functions. The momenta of the incoming parton, $k^\mu$, and 
of the fragmenting parton, $p^\mu$, which appear in the calculation of the partonic cross sections, are approximated 
as $k^\mu\simeq x P^\mu$ and $p^\mu \simeq P_h^\mu /z$, respectively. It is then convenient to work with the partonic 
Mandelstam variables 
\beeq\label{stu}
s=(k+l)^2=xS,\;\, t=(k-p)^2=\frac{x}{z}T,\;\,u=(l-p)^2=\frac{U}{z}\,.
\eeeq
The general form of the factorized cross section for the inclusive hadron production process then is 
\beeq 
E_h \frac{d^3\sigma^{\ell N\to h X}}{d^3P_h}&=&\frac{1}{S}\sum_{i, f}
\int_{0}^1 \frac{dx}{x}\int_{0}^1 \frac{dz}{z^2} \, f^{i/N}(x,\mu) D^{h/f}(z,\mu)\;\hat{\sigma}^{i\to f}(s,t,u,\mu)\;,
\label{invariantcs1}
\eeeq
where $f^{i/N}(x,\mu)$ is the parton distribution function (PDF) for the incoming parton $i$ in the nucleon $N$ 
and $D^{h/f}(z,\mu)$ the corresponding fragmentation function (FF) for parton $f$ fragmenting into hadron $h$, 
both evaluated at a factorization scale $\mu$.  
In Eq.~(\ref{invariantcs1}), $\hat{\sigma}^{i\to f}$ 
is the partonic cross section for the lepton-parton scattering process, $\ell + i\to f+x$, with $x$ an unobserved 
partonic final state. The sum in Eq.~(\ref{invariantcs1}) runs over the different species of partons, quarks, gluons and
antiquarks.  

\begin{figure*}[t]
\centering
\subfloat[]{\includegraphics[width=0.2\textwidth]{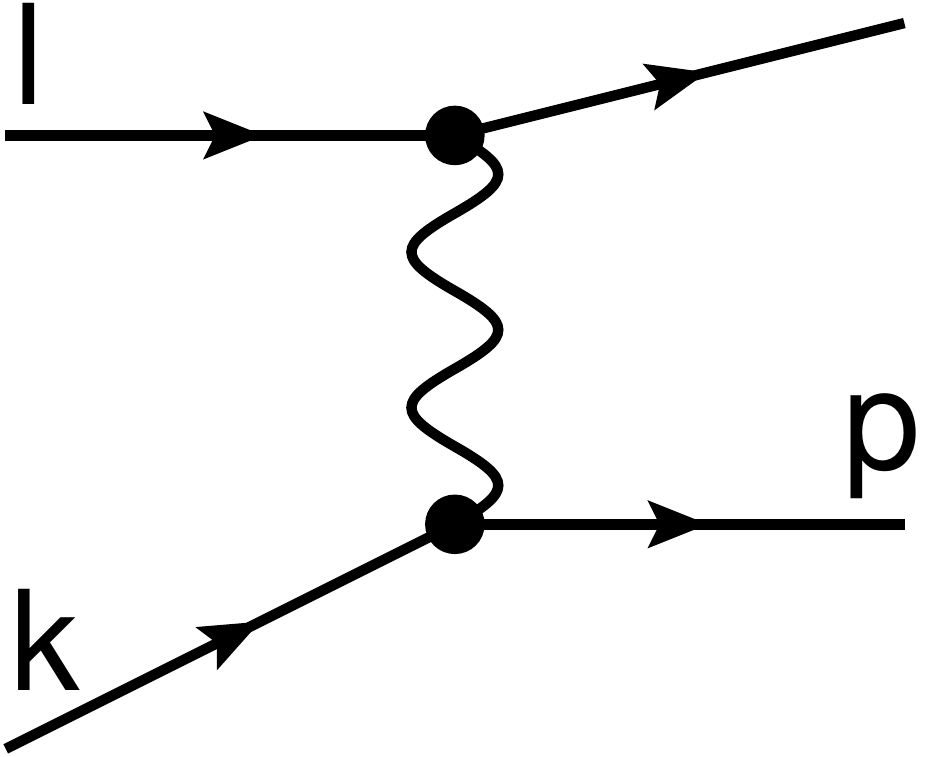}\label{fig:LO}}\hspace{1.5cm}
\subfloat[]{\includegraphics[width=0.6\textwidth]{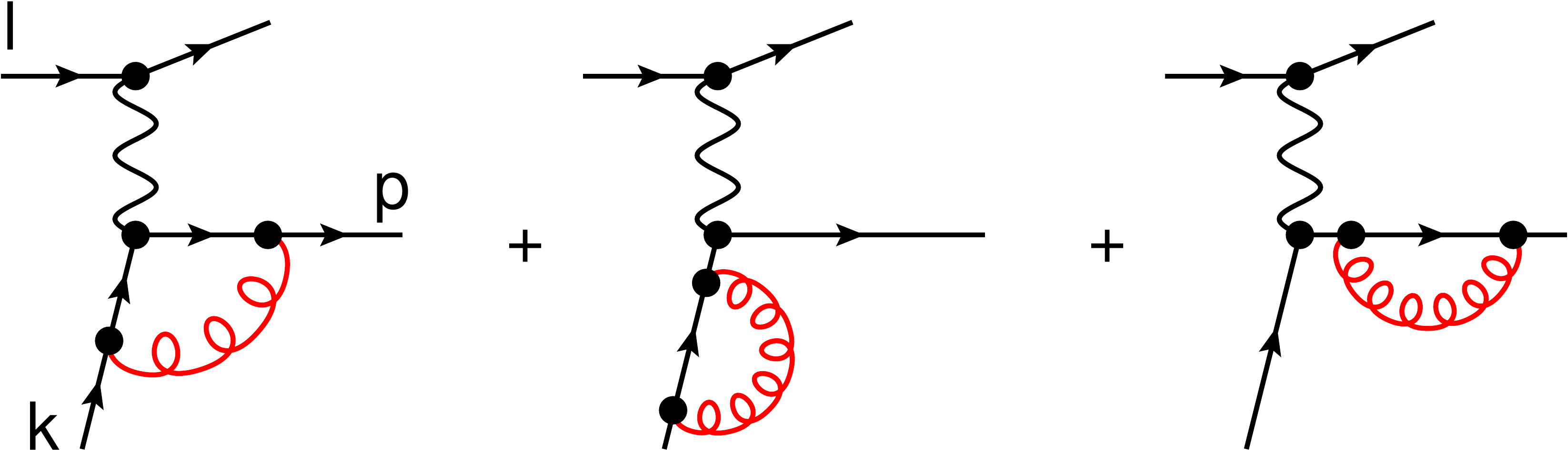}\label{fig:NLOvir}}
\caption{{\bf a)} LO diagram for lepton-quark scattering. {\bf b)} Virtual diagrams at NLO in Feynman gauge.}
\end{figure*}

The partonic cross sections $\hat{\sigma}^{i\to f}$ in Eq.~(\ref{invariantcs1}) can be calculated in QCD perturbation
theory. One may write their expansion in the strong coupling as
\beq
\hat{\sigma}^{i\to f}\,=\,\hat{\sigma}^{i\to f}_{\mathrm{LO}} + \frac{\alpha_s}{\pi}\,\hat{\sigma}^{i\to f}_{\mathrm{NLO}}+
{\cal O}(\alpha_s^2)\,.
\eeq
At lowest order (LO) only the tree-level process $\ell q\to q\ell$ shown in Fig.~\ref{fig:LO} contributes. 
The calculation of its cross section is straightforward. One finds
\beq
\hat{\sigma}_{\mathrm{LO}}^{q\to q}  =  2 \alpha_{\mathrm{em}}^2 e_q^2\;\frac{s^2+u^2}{t^2}\; \delta(s+t+u)\; ,
\label{CSLO}
\eeq
where $\alpha_{\mathrm{em}}$ is the fine structure constant and $e_q$ is the quark's fractional charge.

For a calculation of the hard partonic cross section to NLO accuracy it is convenient to rewrite the $x$- and $z$-integrals in Eq.~(\ref{invariantcs1}) in terms of new 
variables $v=1+t/s$ and $w=-u/(s+t)$. Using~(\ref{stu}), we have
\beq
x=\frac{1-v}{vw}\frac{U}{T}\,,\;\;z=\frac{-T}{(1-v)S}\,,
\eeq
and Eq.~(\ref{invariantcs1}) becomes
\beeq 
E_h \frac{d^3\sigma^{\ell N\to h X}}{d^3P_h}&=&\left(\frac{-U}{S^2}\right)\sum_{i, f}
\int_{\frac{U}{T+U}}^{1+\frac{T}{S}} \frac{dv}{v(1-v)}\int_{\tfrac{1-v}{v}\tfrac{U}{T}}^1 \frac{dw}{w^2} 
 H^{if}(v,w)\;\hat{\sigma}^{i\to f}(v,w,\mu)\;,
\label{Trafox}
\eeeq
where we have defined
\begin{eqnarray}
H^{if}(v,w) & \equiv & \frac{f^{i/N}(x,\mu)}{x}\frac{D^{h/f}(z,\mu)}{z^2} 
\Bigg|_{x=\tfrac{1-v}{vw}\tfrac{U}{T},\,z=\tfrac{-T}{(1-v)S}}\label{SHNx}\,.
\end{eqnarray}
We note that the invariant mass of the unobserved recoiling 
partonic final state is given by $s+t+u=s v (1-w)$. The function $\delta(s+t+u)\propto \delta(1-w)$ in 
the LO cross section~(\ref{CSLO}) expresses the fact that at LO the recoil consists of a single parton.

At the NLO level, the virtual contributions shown in Fig.~\ref{fig:NLOvir} contribute through their interference 
with the Born diagram in Fig.~\ref{fig:LO}. The virtual contributions thus have Born kinematics and are proportional to $\delta(1-w)$. 
Since we are only interested in QCD virtual corrections, only the quark line is affected, and we may adopt
the result directly from the corresponding calculation in Ref.~\cite{Altarelli:1979ub} for the basic photon-quark
scattering diagrams in DIS. This gives
\beeq
\hat{\sigma}^{q\to q}_{\mathrm{NLO,vir}} & = & \frac{C_F \alpha_s(\mu)}{2\pi}  
\frac{\Gamma(1-\varepsilon)^2\Gamma(1+\varepsilon)}{\Gamma(1-2\varepsilon)} \left(\frac{4\pi\mu^2}{-t}\right)^\varepsilon \left(-\frac{2}{\varepsilon^2}-\frac{3}{\varepsilon}-8\right) 
\hat{\sigma}^{q\to q}_{\mathrm{LO},\varepsilon}\;,
\label{virto}
\eeeq
where 
\begin{equation}
\hat{\sigma}^{q\to q}_{\mathrm{LO},\varepsilon} = 
2 \alpha_{\mathrm{em}}^2 e_q^2\frac{1}{sv}\; \Bigg(\frac{1+v^2}{(1-v)^2}-\varepsilon\Bigg)\; \delta(1-w)\; .
\label{CSLOvw1eps}
\end{equation}
is the Born cross section computed in $4-2\varepsilon$ dimensions. Furthermore, $C_F=(N_c^2-1)/2N_c$, 
with $N_c$ the number of colors.

\begin{figure*}[t]
\centering
\subfloat[]{\includegraphics[width=0.25\textwidth]{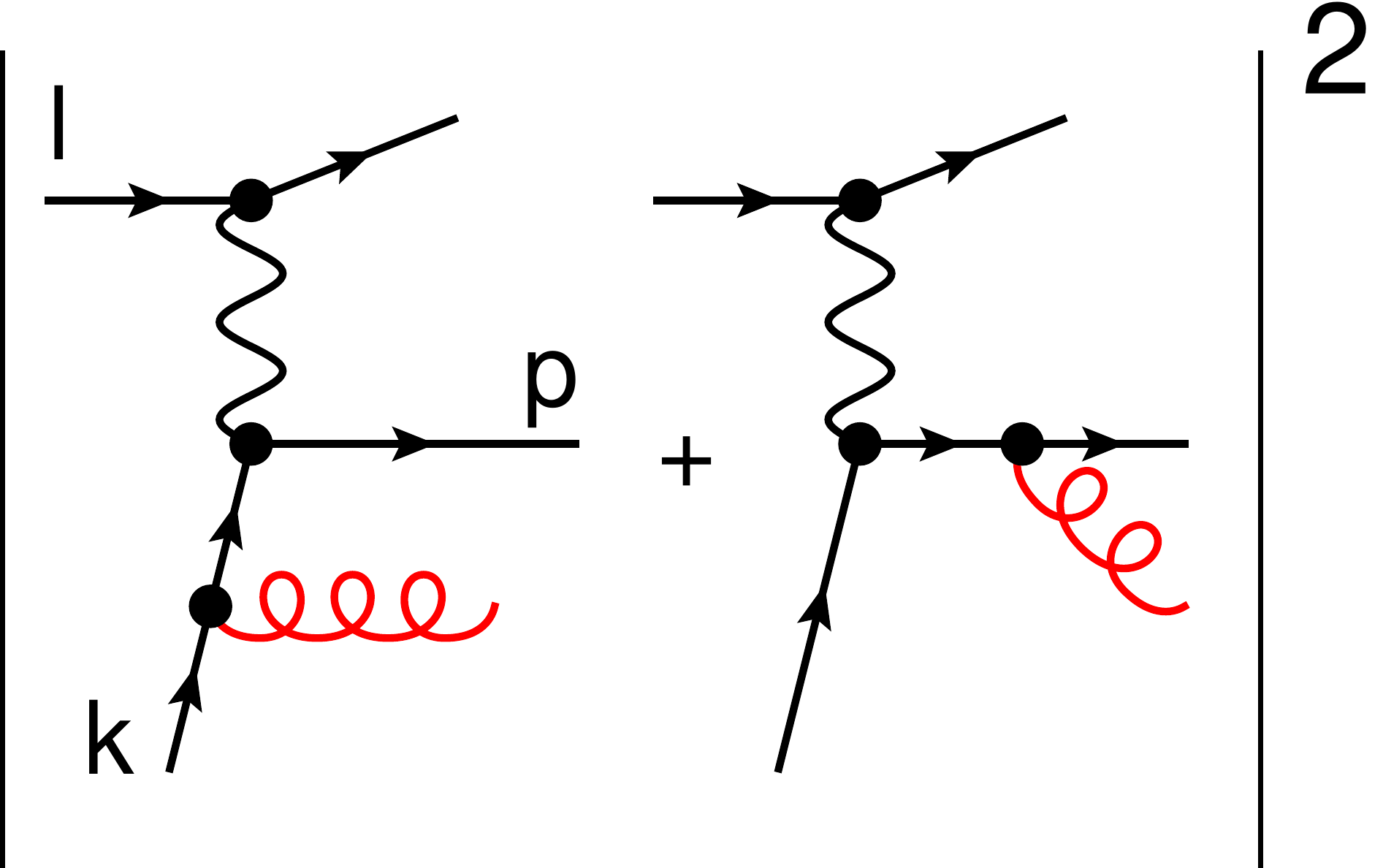}\label{fig:NLOrealq2q}}\hspace{1.5cm}
\subfloat[]{\includegraphics[width=0.25\textwidth]{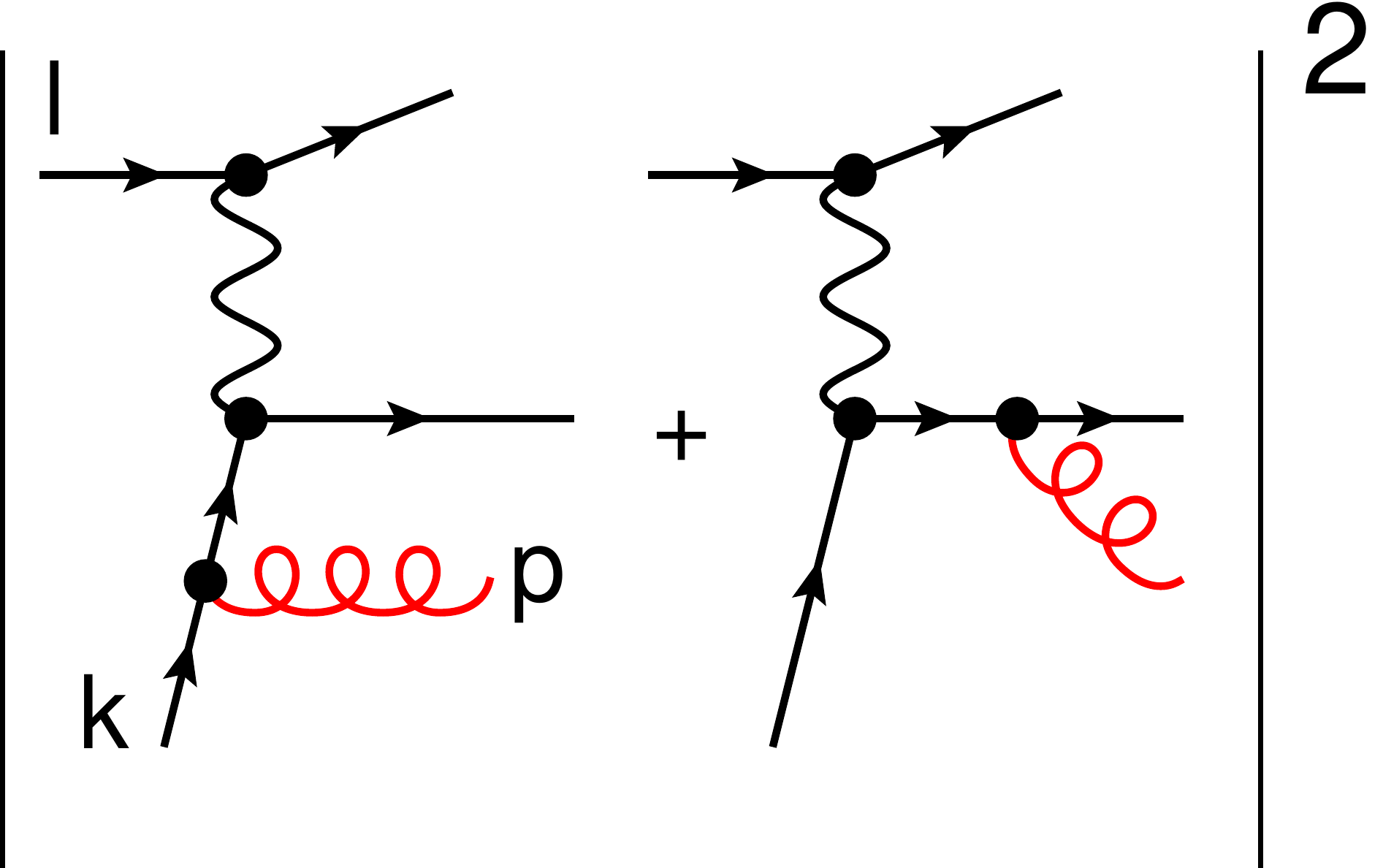}\label{fig:NLOrealq2g}}\hspace{1.5cm}
\subfloat[]{\includegraphics[width=0.25\textwidth]{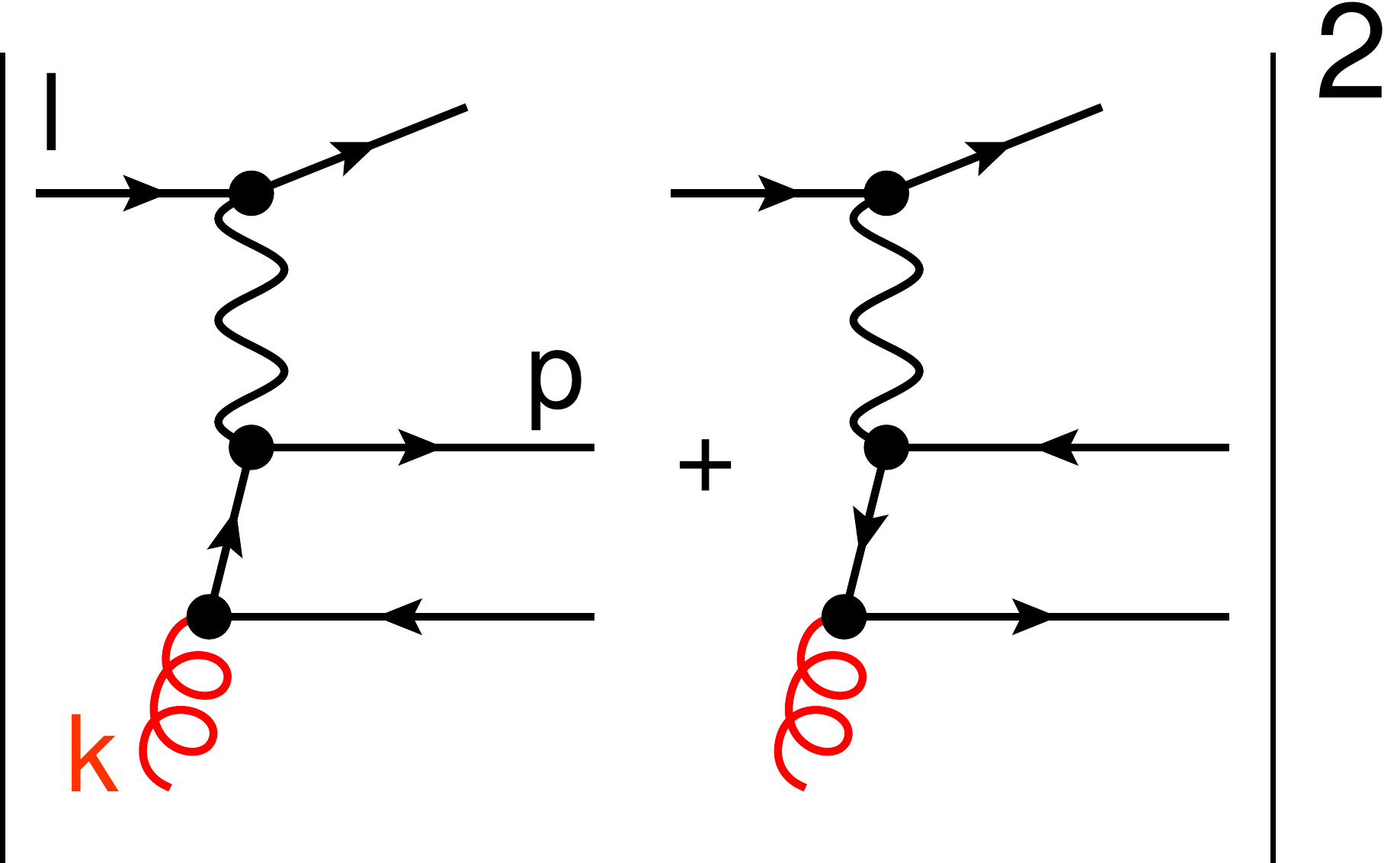}\label{fig:NLOrealg2q}}
\caption{NLO real-emission diagrams. There are three partonic channels at NLO:  (a) $q\to q$, (b) $q\to g$, 
(c) $g\to q$.}
\end{figure*}

The real diagrams in Figs.~\ref{fig:NLOrealq2q},~\ref{fig:NLOrealq2g},~\ref{fig:NLOrealg2q}  have $2\to 3$ topology. To obtain the desired contribution to an inclusive-parton cross
section we need to integrate over the phase space of the lepton and the ``unobserved'' parton in the final state. 
This can be done in $4-2\varepsilon$ dimensions using the standard techniques available in the 
literature~\cite{vanNeerven:1985xr,Beenakker:1988bq,Gordon:1993qc}.

The result for the real NLO contributions in $4-2\varepsilon$ dimensions, although well-defined, is rather lengthy and not presented at this point. The limit $\varepsilon \to 0$ has to be taken with care. In an expansion in $\varepsilon$ one finds a resulting $1/\varepsilon^2$-pole which constitutes itself in the $q\to q$ channel in Fig.~\ref{fig:NLOrealq2q}. This double pole cancels against the $1/\varepsilon^2$-pole of the virtual contribution in Eq.~(\ref{virto}). This behaviour reflects the cancelation of infrared singularities in partonic observables.

After the cancelation of infrared singularities between real and virtual contributions, the 
partonic cross sections still exhibit single poles $1/\varepsilon$ that reflect collinear singularities arising
when an ``observed'' parton (either the incoming one, or the one that fragments) becomes 
collinear with the unobserved parton. The factorization theorem states that these poles may 
be absorbed into the parton distribution functions or into the fragmentation functions. 
This procedure may be formulated in terms of {\it renormalized} parton densities and 
fragmentation functions.

Even after this procedure, one type of collinear singularity remains. It is generated by a momentum configuration 
where the exchanged photon is collinear to the incoming lepton. 
The presence of this singularity is an artifact of neglecting the lepton's mass. 

One approach for dealing with the collinear lepton singularity is to introduce bare and renormalized QED
parton distributions for the lepton $\--$ typically called Weizs\"acker-Williams (WW) distributions \cite{Weizsacker,Williams} $\--$ much in analogy with the renormalization procedure for the nucleon's parton distributions. The only differences are that for leptons the partons are the lepton itself
and the photon, and that we can safely compute their distributions in QED perturbation theory (cf. Ref.~\cite{ourpaper}). 

The hard process involving an incoming lepton will always require two electromagnetic interactions
and hence be of order $\alpha_{\mathrm{em}}^2$, as seen explicitly in Eq.~(\ref{CSLO}). This is different for 
a hard process with an incoming photon such as $\gamma q\to q g$, which is of order $\alpha_{\mathrm{em}}\alpha_s$.
This implies that at NLO in QCD (at order $\alpha_{\mathrm{em}}^2\alpha_s$) there will be contributions
generated by the photon acting as a parton of the lepton and participating in the hard process. 
Such types of contributions are known as Weizs\"acker-Williams contributions.  
In essence, in this situation the lepton merely serves as a source of real photons for those WW - contributions.

Since the Weizs\"acker-Williams distribution is subject to ($\bar{\mathrm{MS}}$)-renormalization as well, the remaining collinear singularity that we encounter in the calculation of the real NLO contributions in Figs.~\ref{fig:NLOrealq2q},~\ref{fig:NLOrealq2g},~\ref{fig:NLOrealg2q} is canceled in this renormalization procedure. Eventually, one ends up with a well-defined NLO result in four dimensions.

One may adopt a second approach to deal with the collinear lepton-photon singularity. In principle one may perform a full calculation in which the lepton's mass is kept finite. This is trivial for the virtual diagrams, since the QCD corrections do not affect
the lepton line. However, inclusion of a lepton mass considerably complicates the phase space integrations for 
the real diagram. Nevertheless, it is possible to compute the relevant integrals using the results given in
Ref.~\cite{Beenakker:1988bq}. One may then expand the result in powers of the lepton mass and neglect 
terms suppressed by powers of $\mathcal{O}(m_\ell)$. In this way, the ``would-be'' collinear singularity
is regularized by the lepton mass and shows up as a term $\sim \ln(m_\ell^2)$. Terms independent 
of $m_\ell$ are also kept. 

One explicitly finds that the two approaches for treating the initial lepton are equivalent:
The full result obtained using the WW contribution agrees with that for $m_\ell\neq 0$,
as long as one only keeps the leading terms.

Next, we briefly present the structure of the final results for the full partonic cross sections in analytic form. Combining the 
cross section~(\ref{Trafox}) for massless leptons with the Weizs\"{a}cker-Williams contribution
we may write the full NLO cross section as
\beeq 
E_h \frac{d^3\sigma^{\ell N\to h X}}{d^3P_h}&=&\left(\frac{-U}{S^2}\right)\sum_{i, f}
\int_{\frac{U}{T+U}}^{1+\frac{T}{S}} \frac{dv}{v(1-v)}\int_{\tfrac{1-v}{v}\tfrac{U}{T}}^1 \frac{dw}{w^2} 
 H^{if}(v,w)\;\left[ \hat{\sigma}^{i\to f}_{\mathrm{LO}} (v)+ 
\frac{\alpha_s(\mu)}{\pi}\,\hat{\sigma}^{i\to f}_{\mathrm{NLO}}(v,w,\mu)\right.\nn\\[2mm]
&&\hspace{3cm}+f^{\gamma/\ell}_{\mathrm{ren}}\left(\tfrac{1-v}{1-vw},\mu\right)\,
\frac{\alpha_s(\mu)}{\pi}\,\hat{\sigma}_{\mathrm{LO}}^{\gamma i\to f}(v,w)
\Bigg]\;,
\label{Trafox1}
\eeeq
where $H^{if}(v,w)$ has been defined in Eq.~(\ref{SHNx}). The LO contribution, present only for the channel 
$q\to q$ with an incoming quark that also fragments, was already given in~(\ref{CSLO}). 
For the NLO term in this channel we find
\begin{eqnarray}
\hat{\sigma}_{\mathrm{NLO}}^{q\to q}(v,w,\mu)&=&\frac{\alpha_{\mathrm{em}}^2 e_q^2C_F}{svw}\Bigg[
A_0^{q\to q} \,\delta(1-w)+A_1^{q\to q} \left(\frac{\ln(1-w)}{1-w}\right)_++ \frac{1}{(1-w)_+}\Bigg\{B_{1}^{q\to q}\ln\left(\frac{1-v}{v(1-v(1-w))}\right)\nonumber\\[2mm]
&+&B_{2}^{q\to q}\ln(1-v(1-w))+B_{3}^{q\to q}\ln\left(\frac{sv^2}{\mu^2}\right)\Bigg\}+C_1^{q\to q}\ln(v(1-w))+C_2^{q\to q}\ln\left(\frac{(1-v)w}{1-vw}\right)\nonumber\\[2mm]
&+& C_3^{q\to q}\ln\left(\frac{1-v}{(1-vw)(1-v(1-w))}\right)+C_4^{q\to q}\ln\left(\frac{s}{\mu^2}\right)+C_5^{q\to q}\Bigg]\,,\label{Resq2qNLOreal1}
\end{eqnarray}
where the coefficients $A_i^{q\to q}$, $B_i^{q\to q}$, $C^{q\to q}_{i}$ are functions of $v$ and $w$ and
their explicit form may be found in Ref.~\cite{ourpaper}. The channels $q\to g$ and $g\to q$ have simpler expressions:
\begin{eqnarray}
\hat{\sigma}_{\mathrm{NLO}}^{q\to g}(v,w,\mu) & = & \frac{\alpha_{\mathrm{em}}^2 e_q^2C_F}{svw}
\Bigg[C_1^{q\to g}\ln(1-v(1-w))+C_2^{q\to g}\ln\left(\frac{1-v}{(1-vw)(1-v(1-w))}\right)\nonumber\\[2mm]
&&\hspace{3cm}+C_3^{q\to g}\ln\left(\frac{v(1-w)s}{\mu^2}\right)+C_4^{q\to g}\Bigg]\,,
\label{Resq2gNLOreal1}\\
\hat{\sigma}_{\mathrm{NLO}}^{g\to q}(v,w,\mu) & = & 
\frac{\alpha_{\mathrm{em}}^2 e_q^2T_R}{svw}
\Bigg[C_1^{g\to q}\ln\left(\frac{(1-v)w}{1-vw}\right)+C_2^{g\to q}\ln\left(\frac{v(1-w)s}{\mu^2}\right)+C_3^{g\to q}\Bigg]\,.
\label{Resg2qNLOreal1}
\end{eqnarray}
The coefficients $C^{q\to g}_{i}$ and $C^{g\to q}_{i}$ are again given in Ref.~\cite{ourpaper}. The partonic cross sections $\hat{\sigma}_{\mathrm{LO}}^{\gamma q\to q}$,  $\hat{\sigma}_{\mathrm{LO}}^{\gamma q\to g}$, $\hat{\sigma}_{\mathrm{LO}}^{\gamma g\to q}$ represent the Weizs\"acker-Williams contributions in Eq. (\ref{Trafox1}) for the relevant channels and we again refer the reader to Ref.~\cite{ourpaper} for their explicit form.

\section{K-factors \label{pheno}}

\begin{figure}[htb]
\centering
\includegraphics[width=0.45\textwidth,angle=270]{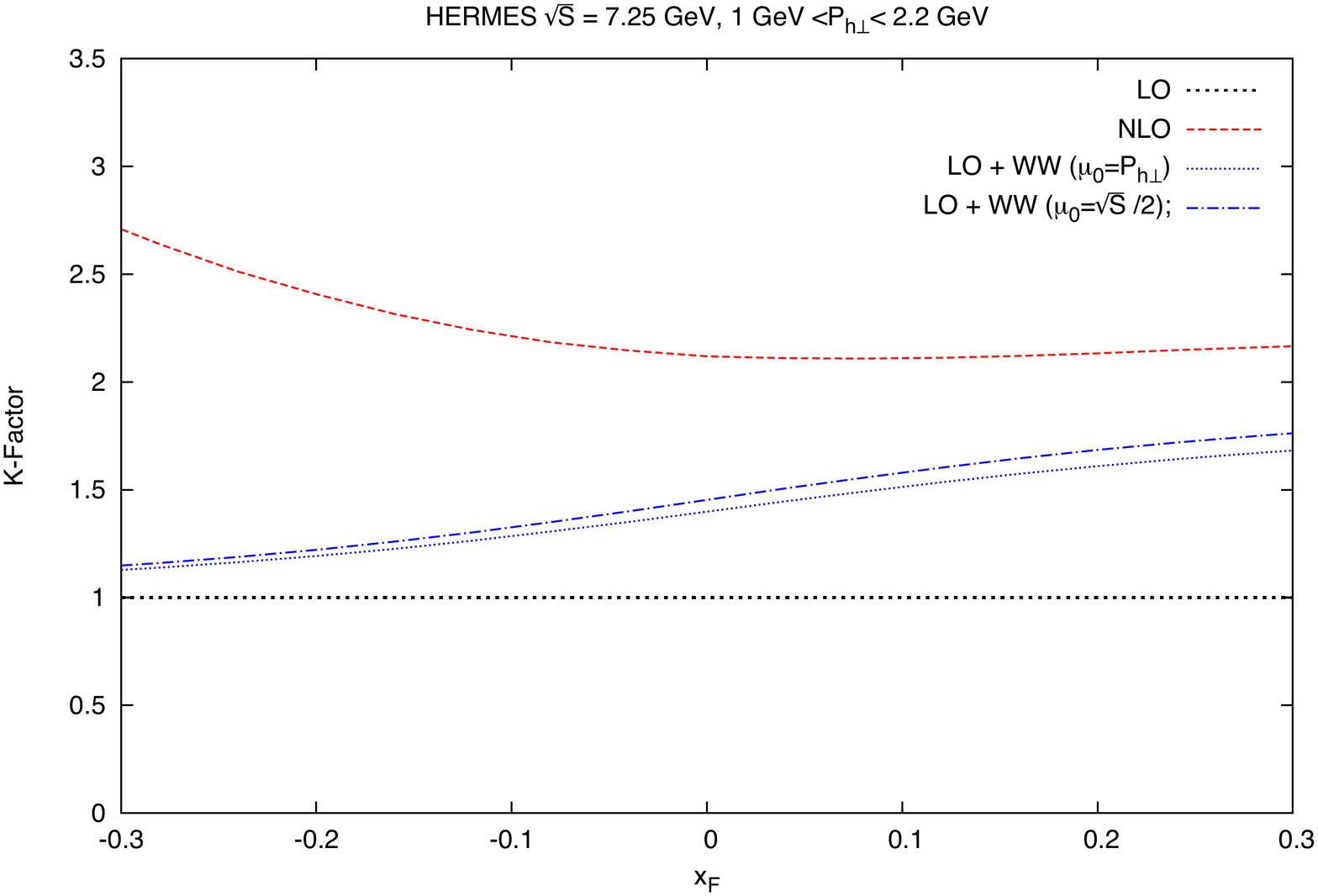}
\caption{K-factor (at a scale $\mu = P_{h\perp}$) for the HERMES experiment plotted vs. the Feynman variable $x_F$ and a binned transverse momentum $P_{h\perp}$.\label{fig:HERMESxf}}
\end{figure}

In Ref.~\cite{ourpaper} we utilized Eq.~(\ref{Trafox1}) in order to give numerical predictions for the spin-averaged cross section of the process $\ell + N\to h + X$ at fixed target experiments (Jefferson Lab, HERMES, COMPASS) and a collider experiment (EIC). We investigated differential cross sections depending on the longitudinal momentum component of the produced hadron (represented by a Feynman variable $x_F$ or pseudorapidity $\eta$) at a  fixed transverse momentum $P_{h\perp}$, or vice versa. We have found particularly large NLO - corrections at low center-of-mass energies of the colliding lepton and nucleon, in particular at JLab and HERMES energies. Although potentially scale dependent, the so-called K-factor is an illustrative quantity to represent the magnitude of the NLO - corrections. It is defined as
\begin{equation}
K_{\mathrm{NLO}}=\frac{\sigma_{\mathrm{NLO}}}{\sigma_{\mathrm{LO}}}.\label{Kfactor}
\end{equation}

\begin{figure}[htb]
\centering
\includegraphics[width=0.45\textwidth,angle=270]{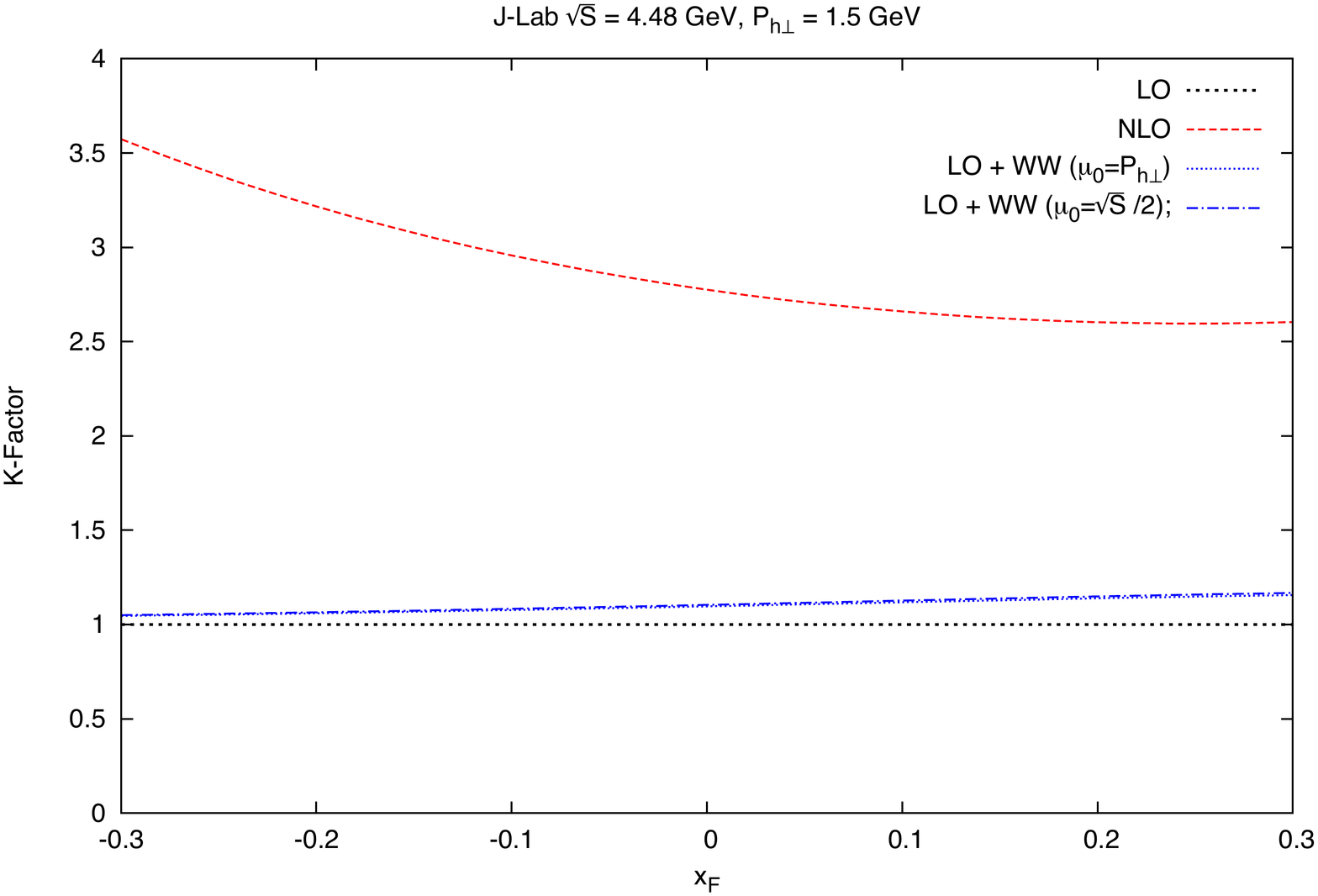}
\caption{K-factor (at a scale $\mu = P_{h\perp}$) for the 12 GeV upgrade at Jefferson Lab plotted vs. $x_F$ and a fixed transverse momentum $P_{h\perp}$.\label{fig:JLabxf}}
\end{figure}

\begin{figure}[htb]
\centering
\includegraphics[width=0.45\textwidth,angle=270]{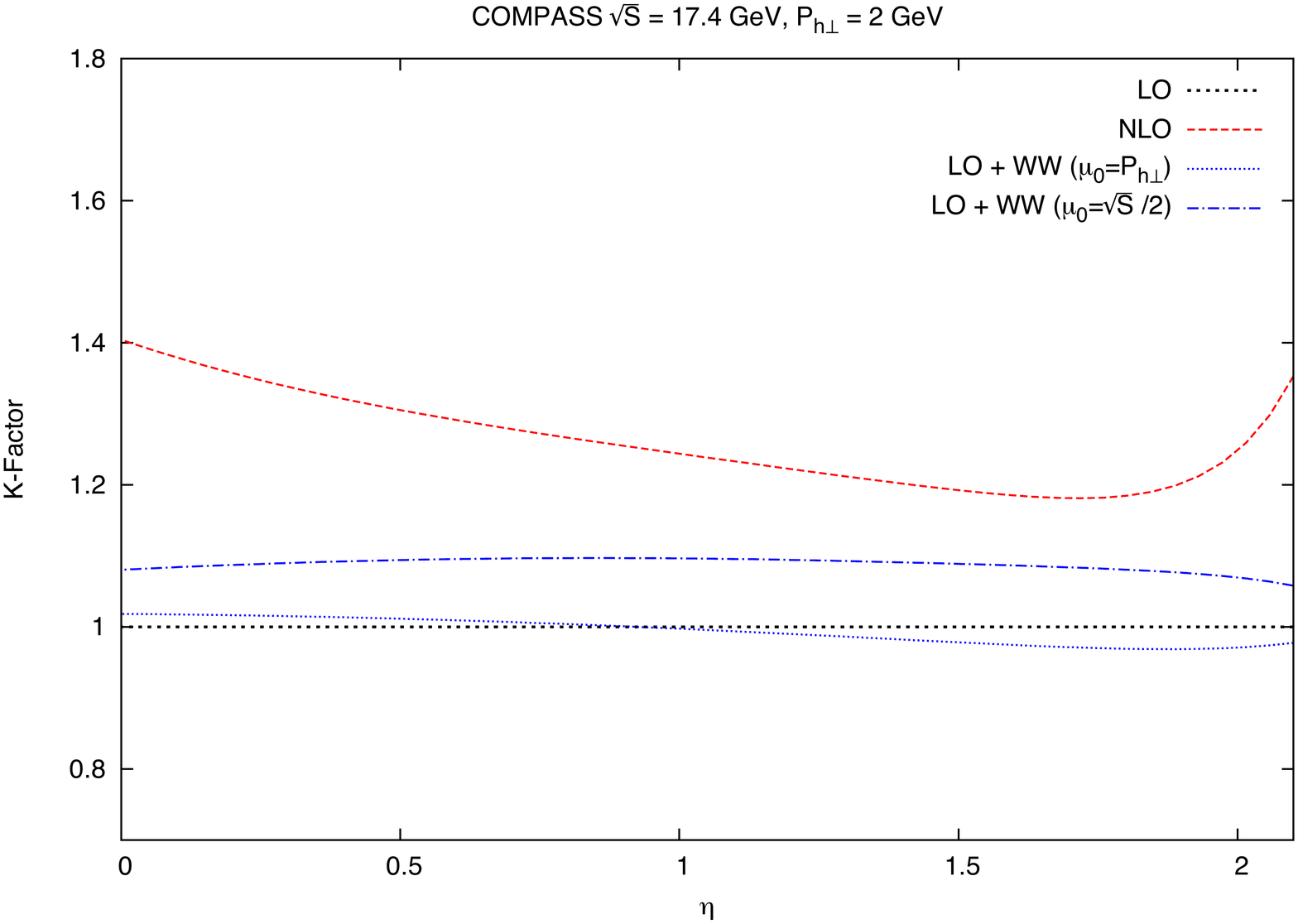}
\caption{K-factor (at a scale $\mu = P_{h\perp}$) for the COMPASS experiment plotted vs. the pseudorapidity $\eta$ and a fixed transverse momentum $P_{h\perp}$.\label{fig:COMPASSeta}}
\end{figure}

In the following we use the numerical results of Ref.~\cite{ourpaper} as input for the K-factors. In Fig.~\ref{fig:HERMESxf} we plot the K-factor for $\pi^+$-production at HERMES at $\sqrt{S}=7.25$~GeV. 
It shows the K-factor as a function of the Feynman variable $x_F$ in a bin
$1\;\mathrm{GeV}<P_{h\perp}<2.2\;\mathrm{GeV}$. 
We also examine the situation where dominance of the Weizs\"acker-Williams contribution is {\it assumed} and the NLO correction caused by $\hat{\sigma}^{i\to f}_{\mathrm{NLO}}$ is {\it assumed} to be negligible (cf. Ref.~\cite{ourpaper}), and plot the corresponding K-factors for two scales $\mu_0=P_{h\perp}$ and $\mu_0=\sqrt{S}/2$ in Fig.~\ref{fig:HERMESxf} ($\mu_0$ indicates a spurious scale in the separation of WW - contributions and partonic NLO - contributions, see Ref.~\cite{ourpaper}). However, Fig.~\ref{fig:HERMESxf} as well as Figs.~\ref{fig:JLabxf} and \ref{fig:COMPASSeta} indicates that the LO term plus the WW contribution alone in Eq.~(\ref{Trafox1}) is not a good approximation for the {\it full} NLO result. We also note that we obtain K-factors of similar size ($\sim 2.5$) in a plot vs. the transverse hadron momentum $P_{h\perp}$. 

In Fig.~\ref{fig:JLabxf} we present our predictions for the K-factor as a function of $x_F$ for $\ell\; {}^3{\mathrm{He}}\to \pi^+X$ in 12~GeV 
scattering at the Jefferson Lab
 where we have assumed a fixed transverse momentum $P_{h\perp}=1.5\;\mathrm{GeV}$. 
We observe in Fig.~\ref{fig:JLabxf} that the NLO corrections are even larger compared to Fig.~\ref{fig:HERMESxf} with a K-factor of about $\sim 2.5 - 3.5$. The $P_{h\perp}$-dependence (not shown here) displays a similar behaviour.

The result of our NLO K-factors for COMPASS kinematics is shown in Figs.~\ref{fig:COMPASSeta}. COMPASS uses a muon beam with energy 160~GeV, resulting in 
$\sqrt{S}=17.4$~GeV. Following the choice made by COMPASS, we use here the c.m.s. pseudorapidity 
$\eta$ of the produced hadron rather than its Feynman-$x_F$. Pseudorapidity is counted as
positive in the forward direction of the incident muon. 
The COMPASS spectrometer roughly covers the region $-0.1<\eta<2.38$. From the $\eta$ dependence 
shown in Fig.~\ref{fig:COMPASSeta} for a fixed transverse momentum $P_{h\perp}=2\;\mathrm{GeV}$ 
we observe that the NLO K-factors are significant but not as large as for HERMES and JLab. Their size is about $\sim 1.2 - 1.4$. Strikingly, the Weizs\"acker-Williams contribution is very small here, even 
for the choice $\mu_0=\sqrt{S}/2$. This may be understood from the fact that the muon mass is about 
200 times larger than the electron mass.

\section{Conclusions\label{concl}}

We have described the next-to-leading order calculations of Ref.~\cite{ourpaper} of the partonic cross sections for the process
$\ell N\to h X$ for which the scattered lepton in the final state is
not detected. In particular we have discussed the situation where the exchanged photon is radiated collinearly to the incoming lepton. We have dealt with this situation in two ways. We have first set the mass to zero and have regularized the ensuing collinear 
singularity in dimensional regularization and then subtracted it by introducing a 
Weizs\"{a}cker-Williams type photon distribution in the lepton.
In the second approach, we have kept the lepton mass in the calculation directly, expanding 
all phase space integrals in such a way that the leading mass dependence is obtained. Both approaches 
give the same result. 

We have presented phenomenological NLO predictions for three experimental setups for the fixed-target 
experiments at HERMES, JLab12 and COMPASS in terms of NLO K-factors, a very illustrative quantity to display the magnitude of NLO-corrections. 
We have found that the K-factors are particularly large for the low energy experiments at JLab12 ($\sqrt{S}=4.48\,\mathrm{GeV}$) and HERMES ($\sqrt{S}=7.25\,\mathrm{GeV}$) where we predict K-factors from $2.5$ up to $3.5$. This behaviour has to be attributed to the fact that for those small energies
the plus distribution terms in Eq.~(\ref{Resq2qNLOreal1}) are large, especially at negative $x_F$ or 
rapidity. For larger energies at COMPASS ($\sqrt{S}=17.4\,\mathrm{GeV}$) the K-factor is $1.2-1.4$. Hence, it is significant but well-behaved.

Because of the large K-factors at HERMES and Jefferson Lab one may wonder whether the perturbative expansion is under control. In particular when analyzing the data for a transversely polarized target \cite{HERMES,Hulse:2015caa,JLab}, one should take care as NLO corrections for transverse spin polarization may be large as well. In fact, we argue that the unpolarized cross section should be measured simultaneously to polarized observables for those experiments in order to ensure that perturbative QCD works for the process $\ell + N\to h+X$ at lower c.m.-energies.


\end{document}